\documentclass{article}
\usepackage{graphicx} 
\usepackage{amsmath}
\usepackage{multirow}
\usepackage{hyperref}
\title{A Two-Step Longstaff Schwartz Monte Carlo Approach to Game Option Pricing}

\author{Ce Wang \\ \href{ryan.wang.19@outlook.com}{ryan.wang.19@outlook.com} 
}
\begin{document}

\maketitle
\begin{abstract}
    We proposed a two-step Longstaff Schwartz Monte Carlo (LSMC) method with two regression models fitted at each time step to price game options. Although the original LSMC can be used to price game options with am enlarged range of path in regression and a modified cashflow updating rule, we identified a drawback of such approach, which motivated us to propose our approach. We implemented numerical examples with benchmarks using binomial tree and numerical PDE, and it showed that our method produces more reliable results comparing to the original LSMC.
\end{abstract}
\section{Introduction}
Game options, also known as Israeli options, proposed by Kifer [1], is an option with characteristics of an American put option, which simutaneously offers the issuer to recall the certificate with a penalty paid to the holder. Kifer [1] showed that in the Black-Scholes market (Black and Scholes [2]) where the prices movement of the underlying asset follows
\begin{equation}
    dS_t = \mu S_t dt + \sigma S_t W_t
\end{equation}
Where $S$ is the stock price, $\mu$, $r$, $\sigma$ are constant stock return, risk-free interest rate, and volatility, and $W_t$ is a standard Brownian motion, the game option has unique non-arbitrage price.\\
\\
The pricing of game option is a difficult problem. Kuhn et al. [3] proposed a simulation framework inspired by Roger [4], and also provided benchmarks via the Canadization method proposed by Carr [5]. The pricing of game options via BSDE has been studied by both Dumitrescu et al. [6], and Essaky and Hassani [7]. Wang and Hu [8] studied the pricing of game options under the Levy market.\\
\\
The penalty paid by the issuer when recalling the option at time $t$, is the pay off that the holder would get if the option is exercised at time $t$, plus a penalty $\delta$. Kifer [9] noted that in the discrete case, the price of a game option, denoted by $G_{i,j}$, where $i$ and $j$ are in the range of stock price and time, can be expressed as
\begin{equation}
    G_{ij} = \min(\text{Exercise} + \delta, \max(\text{Exercise}, \text{Holding}))
\end{equation}
This is very useful when using numerical approaches to price game options. For example, if we use a lattice (binomial tree) or a grid (numerical PDE), the above equation can be directly used to update the node values. The idea can be also used to modify the steps in the Longstaff-Schwartz Monte Carlo method [10], which was originally proposed to price American put options.\\
\\
The Longstaff-Schwartz Monte Carlo method is a simulation framework which address the early-exercise property of the holder. It starts from making Monte Carlo simulation of the underlying stock, and targets to calculate cash flows corresponding to each stock path, under the mechanics of American put options. The option price is estimated by the average of the cash flows discounted to $t=0$. The cash flows are calculated by a backward induction from the maturity date. In order to make early exercise decisions, at each time step we fit a regression model to predict the holding value, which is compared to the exercise value to decide if the option should be exercised early. We noticed that the same approach can be used for the pricing of game options, with a enlarged range of paths considered by the regression step to make sure that it also considers the issuer's option to recall the contract. Furthermore, we proposed a new framework, named a two-step Longstaff Schwartz Monte Carlo method, which gives a more precious range of paths included in the linear regression step, thus increases the quality of the estimation. We also leveraged (2) to implement The binomial tree and numerical PDE approaches, to serve as benchmarks.
\section{Methodology}
In this section, we will firstly discuss how the Langstaff Schwartz Monte Carlo method can be modified to price game options, according to the pay off structure of game options. We will then identity a drawback of such approach, and introduce our 2-step Langstaff Schwartz Monte Carlo method. The performance improvement of our method is presented in numerical examples in section 3. We then explain how to modify the binomial tree and numerical PDE approach to price game options, as they are benchmarks in the numerical examples in section 3.
\subsection{Longstaff - Schwartz Monte Carlo for Game Option}
Longstaff and Schwartz proposed a Monte Carlo framework incorporating linear regression to predict the holding value of an American put options, which is known as the Longstaff Schwartz Monte Carlo method (LSMC). The LSMC method simulates $n$ paths of stock price under the Black-Scholes model, and results $n$ cashflows. The option price is thus the discounted average of the cashflows. At each time step, a linear regression model is build to use stock prices to predict discounted future cashflow, and the predictions are used to make early exercise decision. When American put options are priced, we can clearly identify the paths which are relevant (i.e. the in-the-money paths), and only include those paths in the regression. However, when game options are priced, we have to include all paths, since an out-of-the-money path can have an discounted expected holding value larger than the penalty, and thus relevant to the issuer's option to recall the contract. This is a draw back and is the motivation to our method, which we will discuss in 2.2 with more details. Another modification to note is that when updating the cash flow via backward induction, we update not only the cash flow on the paths where the holder exercises, but also the cash flow where the issuer exercises.\\
\\
Below are the step-to-step details of the Longstaff-Schwartz Monte Carlo method for game option pricing:
\begin{itemize}
    \item Simulate $N$ stock paths under the risk neutral measure of Black-Scholes model from $t\in \{0,...,T\}$
    \item Initialize the corresponding cashflows of the $N$ paths as $CF_i = \max(K-S_{iT}, 0)$ where $i\in \{1,...,N\}$
    \item Fit a linear regression model $\hat{f}$ where the $y$ variable is the discounted cash flow in all paths, and the $x$ variables are the stock prices in all paths, transformed by selected basis functions
    \item Use $\hat{f}$ to predict the holding value on all paths, denoted as $C_i$ 
    \item Identify the paths $N_{\text{Exercise, Holder}}$ where the holder will exercise, i.e. paths where $C_i < K-S_{i, T-1}$.
    \item Identify the paths $N_{\text{Exercise, Issuer}}$ where the issuer will exercise, i.e. paths where $C > K-S_{i, T-1} + \delta$
    \item Update $CF_i$ where $i \in N_{\text{Exercise, Holder}}$ as $CF_i = K - S_{i, T-1}$
    \item Update $CF_i$ where $i \in N_{\text{Exercise, Issuer}}$ as $CF_i = K - S_{i, T-1} + \delta$
    \item Update $CF_i$ were $i \notin N_{\text{Exercise, Holder}} \cup N_{\text{Exercise, Issuer}}$ as $CF_i = e^{-r\Delta t}CF_i$
    \item Repeat the above until we obtain $CF_i$ when $t = 1$
    \item Compute the average of $CF_i$ and multiply by the discount factor $e^{-r\Delta t}$ as the price of the Game option.
\end{itemize}

\subsection{Two-Step Longstaff-Schwartz Monte Carlo for Game Options}
A draw back of the above Longstaff-Schwartz Monte Carlo method is that all paths are used in the regression steps. Although all paths are potentially relevant since the issuer has the option to exercise with penalty, in reality, many of the out-of-the-money paths are eventually not relevant at each $t$, i.e. their discounted expected holding values are smaller than the penalty. Containing these paths in the regression step will impair the regression models fitted at each time step.\\
\\
We propose a two-step Longstaff-schwartz Monte Carlo method to address the above issue. In our method, we firstly use out-of-the-money paths to fit a regression model to predict the cash flow discounted from the next step. We then make a pay off prediction using this regression model, to identify which paths are relevant to the issuer. Then we fit the regression model in the standard Longstaff-Schwartz Monte Carlo method, but use both the in-the-money paths, and the paths we identified using the previous regression model. The rest part of the method is the same with the method we discussed in section 2.1. This approach helps us to exclude mamy out-of-the-money paths which are not relevant, which improves the performance of the second regression model.\\
Below is a step-to-step explaination of the two-step Longstaff-Schwartz Monte Carlo method:
\begin{itemize}
    \item Simulate $N$ stock paths under the risk neutral measure of Black-Scholes model from $t\in \{0,...,T\}$
    \item Initialize the corresponding cash flow of the $N$ paths as $CF_i = \max(K-S_{iT}, 0)$ where $i\in \{1,...,N\}$
    \item Identify the set $N_{\text{out-of-the-money}}$ where the paths are out-of-the-money at $t = T-1$
    \item Fit a linear regression model $\hat{f}_1$ where the $y$ variable is the discounted cash flow from paths in $N_{\text{out-of-the-money}}$, and the $x$ variables are the stock prices from paths in $N_{\text{out-of-the-money}}$, transformed by selected basis functions
    \item Use $\hat{f}_1$ to predict the holding value on the out-of-the-money paths, denoted as $D_i$ 
    \item Identify the set $N_{\text{out-of-sample-relevant}}$ where the holding value is higher than the penalty
    \item Let $N_{\text{in-the-money}}$ be the in-the-money paths. Let $N_{\text{regression}} = N_{\text{in-the-money}} \cup N_{\text{out-of-sample-relevant}}$
    \item Fit a linear regression model $\hat{f}_2$ where the $y$ variable is the discount cash flow from paths in $N_{\text{regression}}$, and the x variables are the stock prices from paths in $N_{\text{regression}}$
    \item Use $\hat{f}_2$ to predict the holding value of paths in $N_{\text{regression}}$, denoted as $C_i$
    \item Identify the paths $N_{\text{Exercise, Issuer}}$ where the issuer will exercise, i.e. paths where $C > K-S_{i, T-1} + \delta$
    \item Update $CF_i$ where $i \in N_{\text{Exercise, Holder}}$ as $CF_i = K - S_{i, T-1}$
    \item Update $CF_i$ where $i \in N_{\text{Exercise, Issuer}}$ as $CF_i = K - S_{i, T-1} + \delta$
    \item Update $CF_i$ were $i \notin N_{\text{Exercise, Holder}} \cup N_{\text{Exercise, Issuer}}$ as $CF_i = e^{-r\Delta t}CF_i$
    \item Repeat the above until we obtain $CF_i$ when $t = 1$
    \item Compute the average of $CF_i$ and multiply by the discount factor $e^{-r\Delta t}$ as the price of the Game option.
\end{itemize}
\subsection{Benchmark Numerical Methods}
\subsubsection{Cox-Ross-Rubinsten Binomial Tree}
In the binomial tree approach of option pricing, a time interval $t \in [0,T]$ is discretized to $N$ steps, and the price of the option and the underlying asset are discretized to $n+1$ nodes at the $n^{th}$ step. The option prices at time $T$ for all stock price nodes can be decided by the terminal option payoff, and the option prices at $t=0$ can thus be determined by backward induction. Denoting $S_{i,j}$ as the stock price on node $j$ at time $t=i$. The stock price movement from $t = i$ to $t = i+1$ are specified by magnitude $u$ and $d$, i.e. $S_{i+1,j} = uS_{i,j}$, and $S_{i+1,j+1} = dS_{i,j}$. $u$ and $d$ need to be chosen so that the limit of the stock tree follows the Black-Scholes model. We also need the risk neutral probability $p$ of the stock moving upwards in order to construct the tree.\\
There are various ways to specify a binomial tree (see Joshi [11]). We implemented the Cox-Ross-Rubinsten tree [12], which has the following specification:
\begin{align}
    \begin{split}
        &u = e^{\sigma\sqrt{\Delta T}}\\
        &d = e^{-\sigma\sqrt{\Delta T}}\\
        &p = \frac{e^{\sigma\sqrt{\Delta T}} - d}{u - d}
    \end{split}
\end{align}
Denoting $G_{T,j}$ as the option price at node $j$ at $t = T$, then $G_{T,j} = \max(K - S_{T,j}, 0)$. For time steps before $T$, the Game option value at a node is the minimum between the exercise value plus the penalty, and the option value of a corresponding American put option, i.e.
\begin{align}
\begin{split}
G_{t,j} &= \min(\text{Exercise + Penalty}, \max(\text{Holding}, \text{Exercise})\\
&=\min(\max(K-S_{t,j},0)+\delta, \max(e^{-r\Delta T}(pG_{t+1,j} + (1-p)G_{t+1,j+1}),\max(K-S_{t,j},0))
\end{split}
\end{align}
\subsubsection{Numerical PDE}
In Black and Scholes [8], the following PDE known as the Black-Scholes PDE for the price of an option $V$ is derived:
\begin{align}
    V_t + \frac{1}{2}\sigma^2 S^2 V_{SS} + rSV_S - rV = 0
\end{align}
With the terminal condition $V_T$ equals the option payoff, depends on the type of the option. In order to make the PDE to be well-posed, we propose the following boundary conditions: $V(S=0, t) = K$, and $\lim_{S\rightarrow \infty} = V(S,t) = 0$. We apply the Crank-Nicolson method to solve the Black-Scholes PDE numerically.\\
When solving the PDE using the finite difference method on a grid where time is discretized with step length $\Delta t$, and stock price is discretized with step length $\Delta S$, and denoting $V_{i,j}$ as the solution of the PDE when $t$ is on the $i^{th}$ grid and stock price is on the $j^{th}$ grid, the Crank-Nicolson [13] method makes the following approximations on the first and second order derivatives in the PDE:
\begin{align}
    \begin{split}
        &\frac{\partial V_{i-\frac{1}{2}, j}}{\partial t} = \frac{V_{i,j} - V_{i-1, j}}{\Delta t}\\
        &\frac{\partial V_{i-\frac{1}{2}, j}}{\partial S} = \frac{1}{2}[\frac{V_{i-1, j+1}-V_{i-1,j-1}}{2\Delta S} + \frac{V_{i, j+1}-V_{i, j-1}}{2\Delta S}]\\
        &\frac{\partial^2 V_{i-\frac{1}{2}, j}}{\partial S^2} = \frac{1}{2}[\frac{V_{i-1, j+1}-2V_{i-1,j}+V_{i-1,j-1}}{\Delta S^2} + \frac{V_{i, j+1}-2V_{i,j}+V_{i, j-1}}{\Delta S^2}]
    \end{split}
\end{align}
Plugging in the above equations into the PDE results in following:
\begin{align}
    \begin{split}
        &-\frac{\Delta t}{4}(\sigma^2 j^2 - rj)V_{i-1, j-1} + (1+\frac{\Delta t}{2}(\sigma^2 j^2 + r))V_{i-1, j} - \frac{\Delta t}{4}(\sigma^2 j^2 + rj)V_{i-1, j
        +1}\\
        =&\frac{\Delta t}{4}(\sigma^2 j^2 - rj)V_{i, j-1} + (1-\frac{\Delta t}{2}(\sigma^2 j^2 + r))V_{i, j} + \frac{\Delta t}{4}(\sigma^2 j^2 + rj)V_{i, j
        +1}
    \end{split}
\end{align}
Suppose the stock price is divided into grid points $0,...,N_{\text{grid}}$. Then $V$ is specified on $i=0$ and $i = N_{\text{grid}}$ by the boundary condition, and $V$ on $i \in {1,...,N_{\text{grid}-1}}$ can be iteratively solved by the terminal condition. In particular, when Game option is priced, we have:
\begin{align}
\begin{split}
G_{t,j} &= \min(\text{Exercise + Penalty}, \max(\text{Holding}, \text{Exercise})\\
&=\min(\max(K-S_{i,j},0)+\delta, \max(V_{i,j},\max(K-S_{i,j},0)))
\end{split}
\end{align}

\section{Numerical Examples}
In this sections, we present some numerical examples of using the Longstaff-Schwartz Monte Carlo, two-step Longstaff-Schwartz Monte Carlo, Cox-Ross-Robinsten binomial tree, and the numerical PDE to price game options. The purpose is to examine if the performance of the two-step Longstaff-schwartz Monte Carlo is better from the Longstaff-Schwartz Monte Carlo without the second regression step, and use the binomial tree and numerical PDE approaches as benchmarks.\\
\\
\begin{tabular}{ |p{3cm}||p{2cm}|p{2cm}|p{2cm}|p{2cm}| }
 \hline
 \multicolumn{5}{|c|}{$\delta = 10$, $S_0 = 100$, $r = 0.03$, $T = 1$} \\
 \hline
 Other Parameters& LSMC &Two-Step LSMC&Binomial Tree&Numerical PDE\\
 \hline
 $\sigma = 0.2$, $K = 100$   & 6.52313    &6.75100&   6.74290  & 6.74006\\
 $\sigma = 0.2$, $K = 110$&   12.55097  & 12.73824   &12.72614  &12.72369\\
 $\sigma = 0.2$, $K = 90$ &2.67519 & 2.86182&  2.86320    &2.86109\\
$\sigma = 0.25$, $K = 100$&8.37909 & 8.66110&  8.67473 &8.67252\\
 $\sigma = 0.25$, $K = 110$&14.36542  & 14.57393&14.57888    &14.57688\\
 $\sigma = 0.25$, $K = 90$& 4.20261  & 4.44813   &4.42916 &4.42726\\
$\sigma = 0.15$, $K = 100$& 4.63753  & 4.82596&4.82058  &4.81675\\
$\sigma = 0.15$, $K = 110$& 11.91566  & 11.02133&11.04919  &11.04612\\
$\sigma = 0.15$, $K = 90$& 1.35713  & 1.46590&1.46258  &1.46014\\
 \hline
\end{tabular}
\\
\\
\\
\\
\begin{tabular}{ |p{3cm}||p{2cm}|p{2cm}|p{2cm}|p{2cm}| }
 \hline
 \multicolumn{5}{|c|}{$\delta = 5$, $S_0 = 100$, $r = 0.03$, $T = 1$} \\
 \hline
 Other Parameters& LSMC &Two-Step LSMC&Binomial Tree&Numerical PDE\\
 \hline
 $\sigma = 0.2$, $K = 100$   & 4.63699    &4.81565&   4.82058  & 4.81675\\
 $\sigma = 0.2$, $K = 110$&   7.42106  & 7.55843   &7.56807  &7.56345\\
 $\sigma = 0.2$, $K = 90$ &2.65930 & 2.79413&  2.80641    &2.80303\\
$\sigma = 0.25$, $K = 100$&2.78096 & 2.91380&  2.92587 &2.92006\\
 $\sigma = 0.25$, $K = 110$&5.69741  & 5.80393&5.80730    &5.79815\\
 $\sigma = 0.25$, $K = 90$& 1.14396  & 1.23007   &1.22526 &1.22093\\
$\sigma = 0.15$, $K = 100$& 3.73737  & 3.87273&3.86775  &3.86313\\
$\sigma = 0.15$, $K = 110$& 6.52621  & 6.65710&6.65279  &6.64865\\
$\sigma = 0.15$, $K = 90$& 1.89164 & 1.99905&1.98590  &1.98217\\
 \hline
\end{tabular}
\\
\\
From the above examples, it is clear that our two benchmarks via Cox-Ross-Robinsten binomial tree and numerical PDE reconciles well, and can be considered as the 'true value' of game options. Results from our two-step Longstaff-schwartz Monte Carlo method are significantly closer to the benchmarks comparing to the Longstaff-schwartz Monte Carlo method without the extra regression step that we propose. Therefore, we are confident to conclude that the extra regression step we propose improves the performance of the Longstaff-Schwartz Monte Carlo method for pricing game options, and produces trustworthy results.

\end{document}